\documentclass[aip,jcp,reprint,amsmath,amssymb]{revtex4-1}
\usepackage{graphicx}
\usepackage{dcolumn}
\usepackage{bm}
\usepackage[utf8]{inputenc}
\usepackage[T1]{fontenc}
\usepackage{mathptmx}
\usepackage{etoolbox,float}
\usepackage{booktabs}
\usepackage[version=3]{mhchem}

\makeatletter
\def\@email#1#2{%
 \endgroup
 \patchcmd{\titleblock@produce}
  {\frontmatter@RRAPformat}
  {\frontmatter@RRAPformat{\produce@RRAP{*#1\href{mailto:#2}{#2}}}\frontmatter@RRAPformat}
  {}{}
}%
\makeatother

\begin{document}
\preprint{}

\title{Surface diffusion of phosphorus on Si(100) after PBr$_3$ adsorption}

\author{T. V. Pavlova}
 \email{pavlova@kapella.gpi.ru}
\affiliation{Prokhorov General Physics Institute of the Russian Academy of Sciences, Vavilov str. 38, 119991 Moscow, Russia}
\affiliation{HSE University, Myasnitskaya str. 20, 101000 Moscow, Russia}

\author{V. M. Shevlyuga}
\affiliation{Prokhorov General Physics Institute of the Russian Academy of Sciences, Vavilov str. 38, 119991 Moscow, Russia}

\date{\today}

\begin{abstract}

Phosphorus diffusion on a Si(100) surface was studied using scanning tunneling microscopy (STM) at temperatures of 77 and 300\,K. The phosphorus source utilized was the PBr$_3$ molecule, which fully dissociates on the surface at 77\,K. We observed diffusion of P atoms both along and across the rows of Si dimers. To support the observation of different diffusion pathways of phosphorus, activation energy calculations were performed using density functional theory. At 77\,K, phosphorus diffusion started and (or) finished mostly in bridge positions. At 300\,K, phosphorus diffuses predominantly between end-bridge positions, accompanied by bromine diffusion. The presence of Br near phosphorus significantly restricts its mobility. Additionally, phosphorus was found to diffuse to an oxygen atom that appeared on the surface as a result of water adsorption. This diffusion occurs because the P site near the oxidized dimer is more stable compared to that on the clean surface. The obtained results complement the knowledge about the interaction of phosphorus with the silicon surface, specifically the phosphorus diffusion pathways on the Si(100) surface.

\end{abstract}

\maketitle

\section{Introduction}

Phosphorus diffusion can occur during the technologically important process of silicon doping for microelectronics. In the case of silicon doping by chemical vapor deposition, phosphine is typically used as a phosphorus source gas. Phosphine is dissociatively adsorbed on the S(100) surface at room temperature \cite{2016Warschkow}, and the phosphorus is incorporated into silicon upon heating \cite{2003Schofield}. Before being incorporated into silicon, phosphorus is able to move across the surface \cite{2016Warschkow}, which makes diffusion an important part of the phosphorus-surface interaction. The study of P diffusion on the silicon surface contributes to a better understanding of phosphorus distribution in the doping process. For atomically precise doping \cite{2003Schofield, 2022Wyrick}, study of phosphorus diffusion is particularly important in order to minimize undesired movement, since a dopant shift even by one Si lattice constant is crucial \cite{2023Jones, 2024Hsueh}.

For atomically precise doping of silicon using the PH$_3$ molecule, the mobility of the molecular fragments on the Si(100) surface after adsorption and before phosphorus incorporation was investigated at room temperature. Notably, the fragment PH$_2$ was found to be very mobile \cite{2006Reusch, 2006SchofieldJPCB, 2016Warschkow}. It was also found that the P atom moved along a dimer row between the most favorable end-bridge positions \cite{2016Warschkow}. The phosphorus diffusion was calculated earlier for paths along and across Si dimer rows \cite{1992Brocks} and between the end-bridge and bridge positions \cite{2009Bennett}. The observed phosphorus diffusion between the end-bridge positions was suggested to follow the pathway passing via the bridge site \cite{2016Warschkow}. To our knowledge, no other P diffusion paths on Si(100) have been observed or calculated.

The current study focuses on different P diffusion paths on the Si(100) surface along and across dimer rows at temperatures of 77\,K and 300\,K. Phosphorus diffusion was investigated using a scanning tunneling microscope (STM) in conjunction with density functional theory (DFT) calculations. We chose PBr$_3$ as the phosphorus source gas because the PBr$_3$ molecule completely dissociates to a single phosphorus atom after adsorption on Si(100)at 300\,K \cite{2023Shevlyuga}. Moreover, adsorption positions of all atoms after PBr$_3$  dissociative adsorption are well known \cite{2023Shevlyuga}. Additional interest in the PBr$_3$ molecule arises from its potential application as an alternate source of phosphorus for atomically precise Si doping \cite{2023Shevlyuga, 2024Pavlova, 2025NanoFutures}. Different diffusion paths of phosphorus atoms were observed and the corresponding minimum energy pathways were calculated. In particular, we directly observed the diffusion path between the end-bridge and the bridge sites, calculated previously \cite{1992Brocks, 2009Bennett}. The presence of bromine was shown to significantly inhibit the mobility of phosphorus. Additionally, phosphorus diffusion near defects on Si(100), which are oxygen atoms, was observed.

\begin{figure}[h]
\begin{center}
\includegraphics[width=\linewidth]{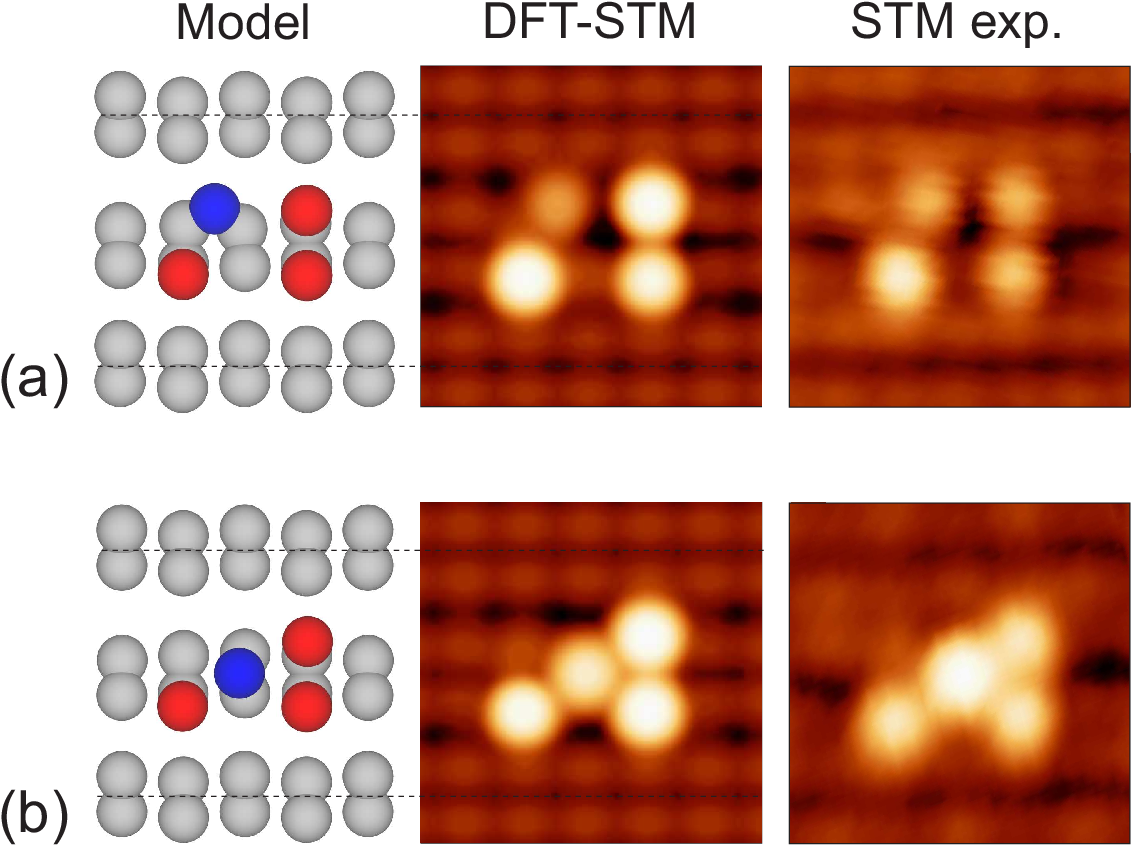}
\caption{\label{fig1} Model, simulated and experimental empty state STM images of structure S1 (a) and S2 (b) after PBr$_3$ dissociative adsorption on the Si(100) surface. Si atoms are shown in gray, Br in red, and P in blue. The dashed line denotes the middle of the dimer row. In the most stable structure, S1 (a), the P atom occupies an end-bridge position between two adjacent dimers and above the Si atom of the previous layer. In structure S2 (b), the P atom occupies a bridge position on the dimer, slightly displaced from its center. The Br atoms are located above the Si atoms slightly away from the center of the dimer. Experimental empty state STM images ($U_s =+2.3$\,V, I$_t$ = 2.0\,nA) were recorded at 77\,K. The voltage ($U_s$) was applied to the sample.}
\end{center}
\end{figure}

\begin{figure}[h]
\begin{center}
\includegraphics[width=\linewidth]{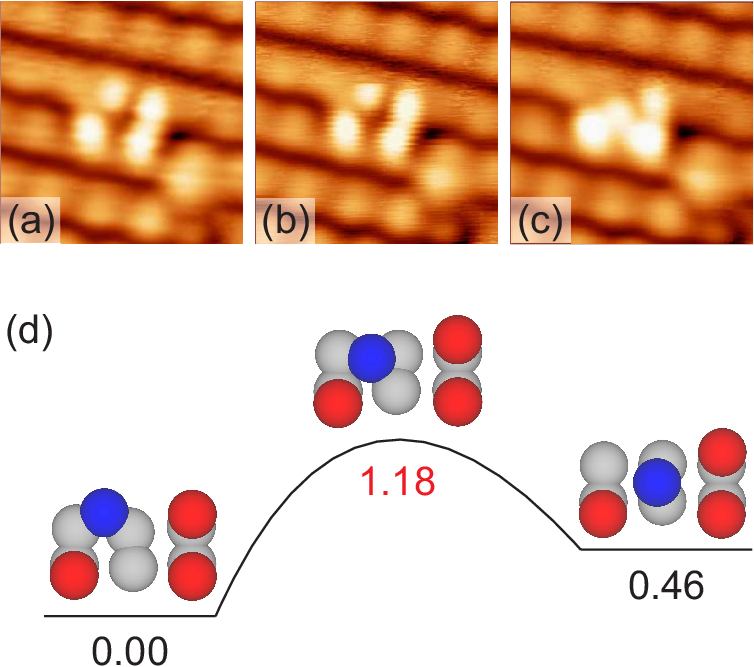}
\caption{\label{fig2} Phosphorus diffusion from the end-bridge to bridge position on Si(100). (a--c) Empty state STM images ($U_s =+2.5$\,V, I$_t$ = 1.0\,nA, 77\,K) of structure S1 (a), S1 with a line scan break above the P atom indicating transition from the end-bridge to the bridge position (b), and structure S2 (c). Scanning was performed from the bottom to top. In the empty state STM images, the dark stripes along the dimer rows pass through the center of the dimer. (d) Energy barrier diagram of the P diffusion from the end-bridge to the bridge site. All values are given in electronvolts, the red number denotes the activation barrier. Si atoms are marked in gray, Br atoms in red, and the P atom in blue.}
\end{center}
\end{figure}

\begin{figure}[h]
\begin{center}
\includegraphics[width=\linewidth]{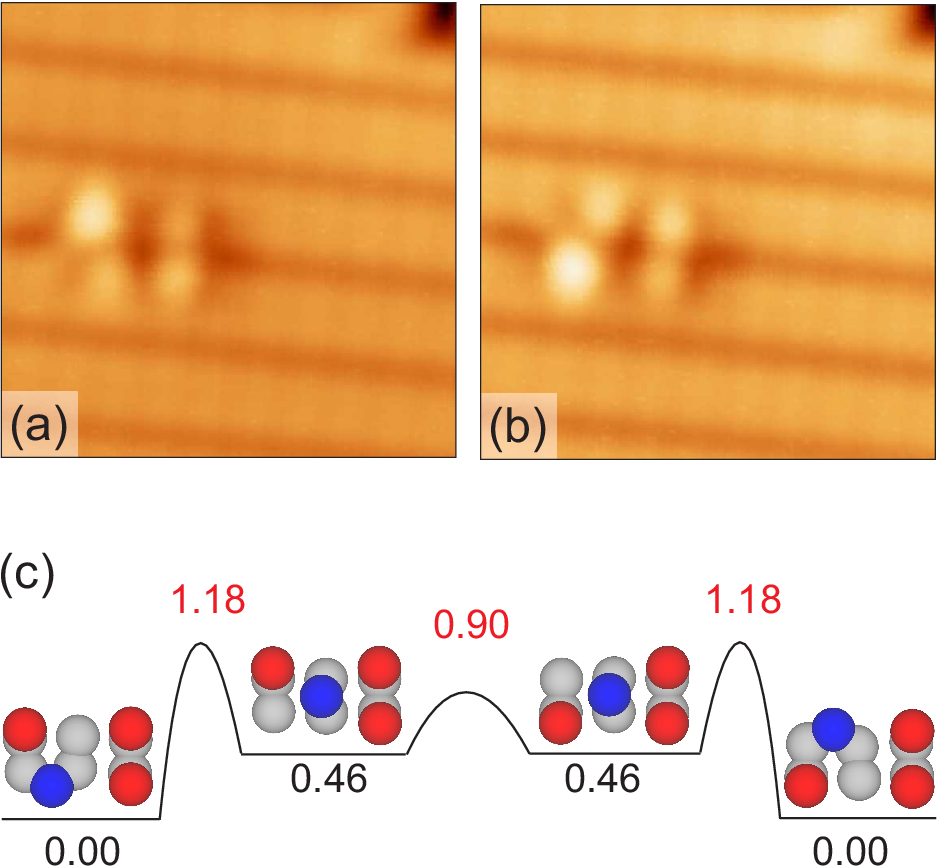}
\caption{\label{fig3} Phosphorus diffusion between end-bridge positions on the opposite sides of a dimer row. Empty state STM images ($U_s =+1.5$\,V, I$_t$ = 2.5\,nA, 300\,K) of the initial (a) and final surface structure (b). (c) Energy barrier diagram for the P diffusion between two end-bridge positions across the dimer row. All values are given in electronvolts, the red numbers denote the activation barriers. Si atoms are marked in gray, Br atoms in red, and the P atom in blue.}
\end{center}
\end{figure}

\begin{figure}[h]
\begin{center}
\includegraphics[width=\linewidth]{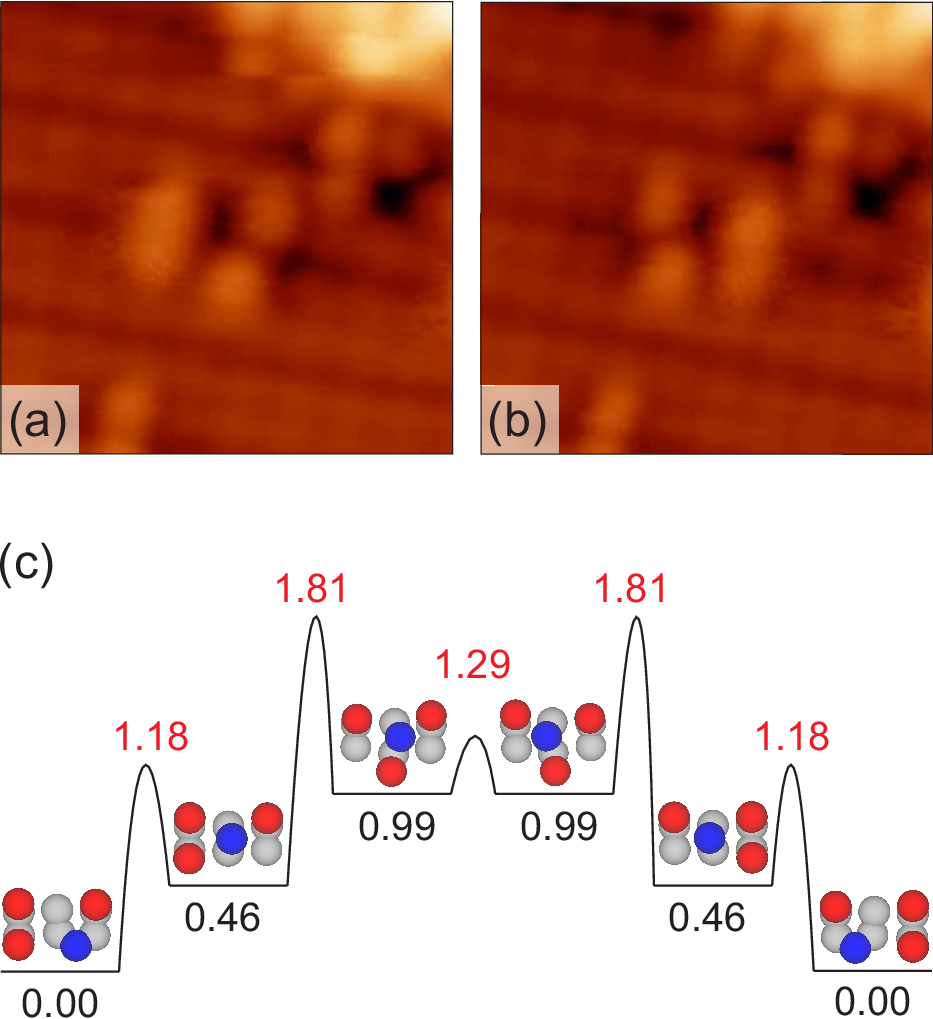}
\caption{\label{fig4} Phosphorus diffusion between end-bridge positions on the same side of a dimer row. Empty state STM images ($U_s =+1.5$\,V, I$_t$ = 2.5\,nA, 300\,K) of the initial (a) and final surface structure (b). (c) Energy barrier diagram for the P diffusion between two end-bridge positions within the dimer row through the bridge position on the central dimer. All values are given in electronvolts, the red numbers denote the activation barriers. Si atoms are marked in gray, Br atoms in red, and the P atom in blue.}
\end{center}
\end{figure}

\begin{figure*}[t]
\begin{center}
\includegraphics[width=0.8\linewidth]{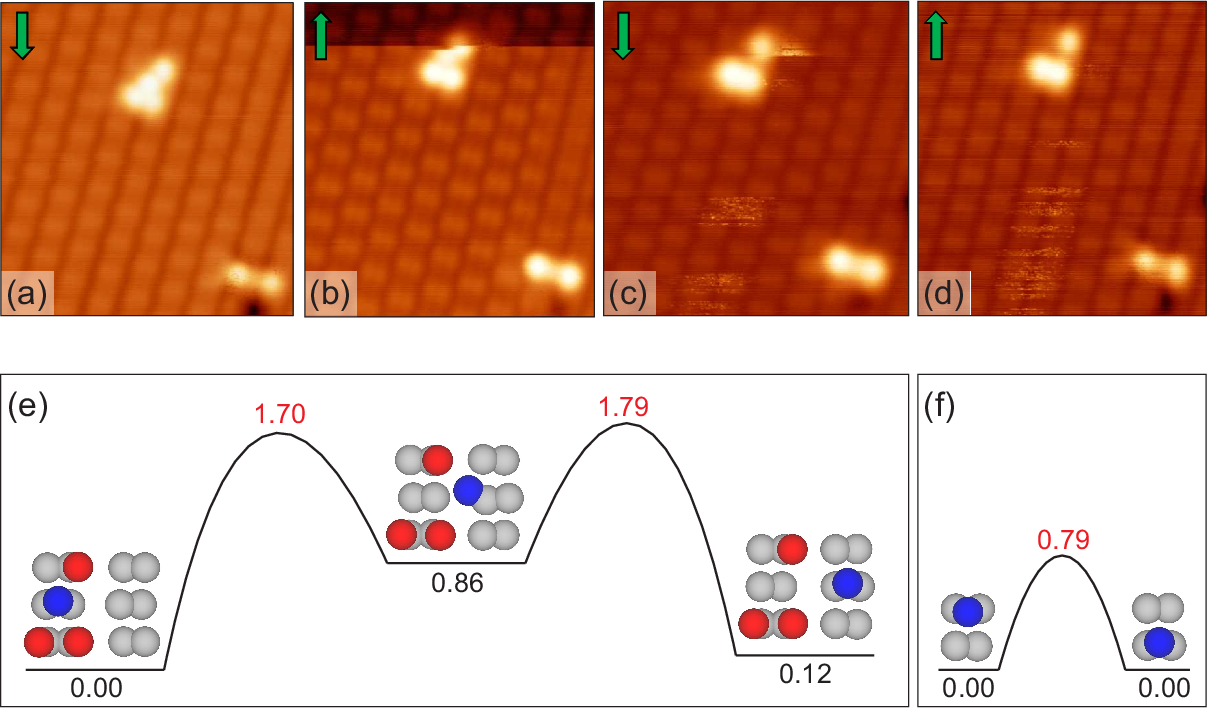}
\caption{\label{fig5} (a--d) Empty state STM images ($U_s =+3.0$\,V, I$_t$ = 2.0\,nA, 77\,K) of P diffusion between the bridge positions. Initially P was located in structure S2 (a), then jumped to the adjacent dimer row (b) and moved within the row (c,d). Arrows show the scanning direction. (e) Energy barrier diagram of the P diffusion between dimer rows. (f) Energy barrier diagram of the P diffusion within the dimer row. All values are given in electronvolts, the red numbers denote the activation barriers. Si atoms are marked in gray, Br atoms in red, and the P atom in blue.}
\end{center}
\end{figure*}

\begin{figure*}[t!]
\begin{center}
\includegraphics[width=0.8\linewidth]{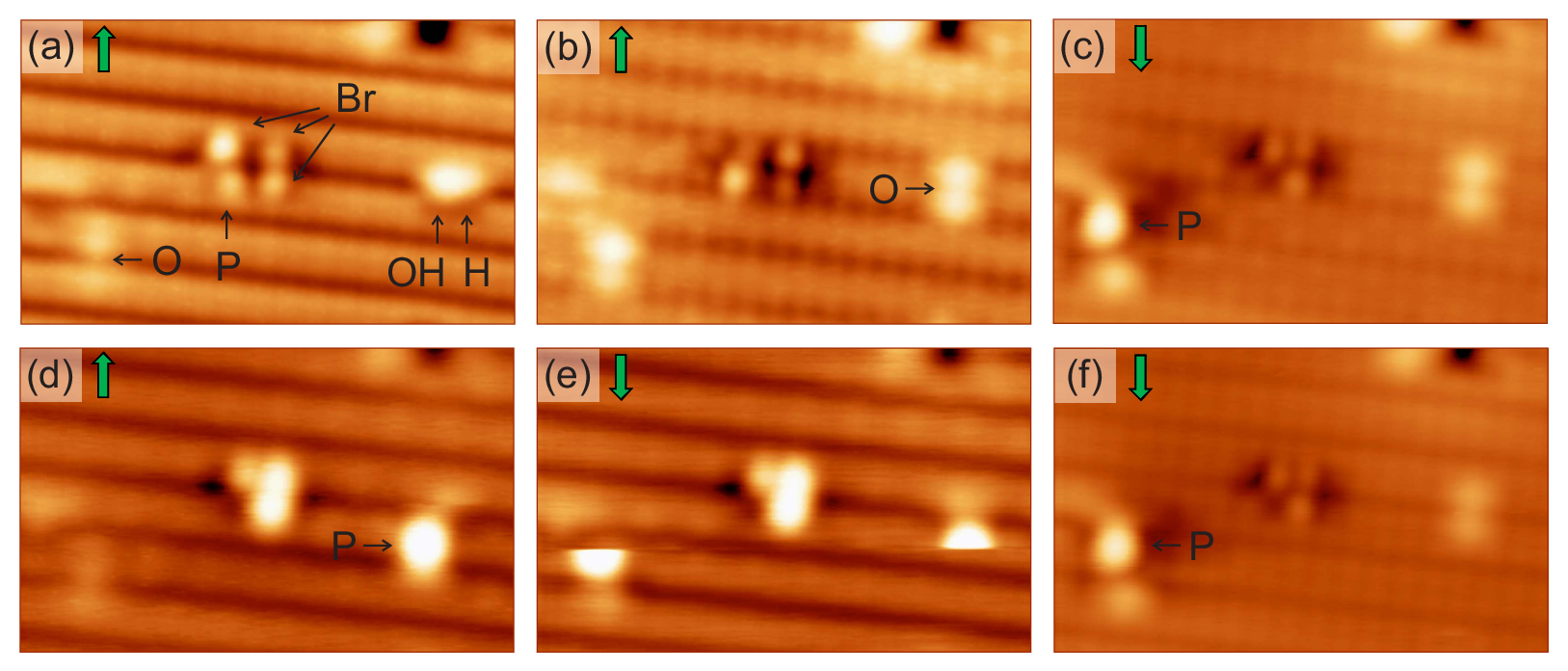}
\caption{\label{fig6} Diffusion of phosphorus between positions adjacent to oxidized dimers. A series of empty state STM images recorded at 300\,K: (a) $U_s =+2.7$\,V, I$_t$ = 2.3\,nA, (b) $U_s =+1.0$\,V, I$_t$ = 1.9\,nA, (c) $U_s =+1.1$\,V, I$_t$ = 1.0\,nA, (d) $U_s =+2.2$\,V, I$_t$ = 1.0\,nA, (e) $U_s =+2.2$\,V, I$_t$ = 1.0\,nA, (f) $U_s =+1.2$\,V, I$_t$ = 1.0\,nA. Arrows show the scanning direction. Phosphorus diffuses between two end-bridge positions on the opposite sides of a dimer row in (a) and (b), moves to oxygen in (c), moves to the other oxidized dimer in (d), and hops back in (e,f).}
\end{center}
\end{figure*}

\begin{figure}[t!]
\begin{center}
\includegraphics[width=0.8\linewidth]{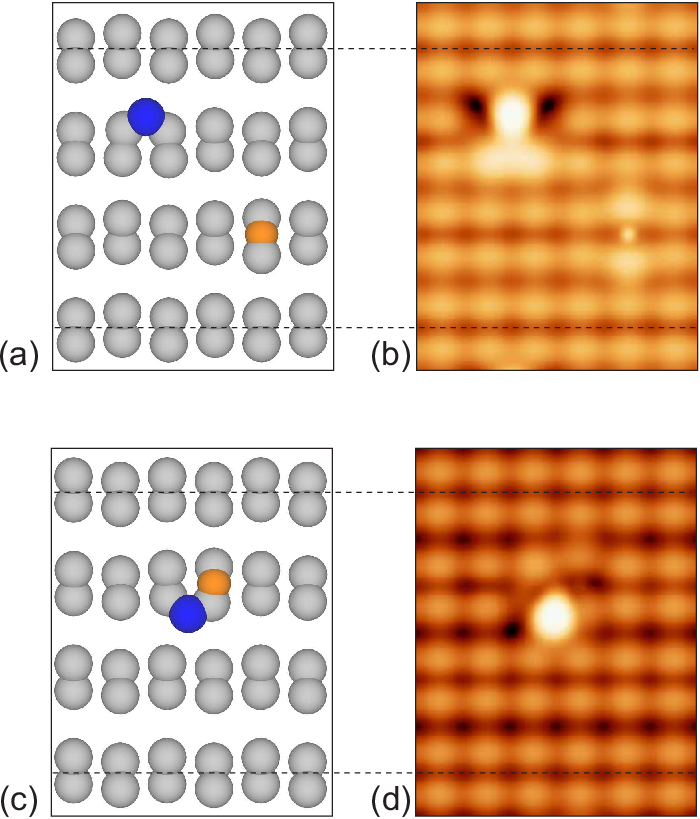}
\caption{\label{fig7} Co-adsorption of phosphorus and oxygen on Si(100). Model and simulated empty state STM image of phosphorus and oxygen atoms located in different dimer rows (a,b) and close to each other (c,d). Si atoms are shown in grey, P in blue and O in orange. The dashed line denotes the middle of the dimer row.}
\end{center}
\end{figure}

\section{Experimental and computational details}

An ultra-high vacuum (UHV) system with a base pressure of 5$\times$10$^{-11}$\,Torr was used for the experiments. The STM measurements were carried out in GPI CRYO (SigmaScan Ltd.) at 77 and 300\,K. B-doped Si(100) samples (1\,$\Omega$\,cm) were prepared by outgassing the wafer at 1170\,K overnight in UHV followed by flash-annealing at 1470\,K. The \ce{PBr3} adsorption on Si(100) was carried out at 77 and 300\,K at a partial pressure of $3 \cdot 10^{-10}$\,Torr for one minute. We used mechanically cut Pt-Rh and Pt-Ir tips, as well as polycrystalline W tips that had been electrochemically etched. To process STM images, the WSXM program \cite{WSXM} was utilized.

The spin-polarized DFT calculations were performed in the Vienna \textit{ab initio} simulation package (VASP) \cite{1996Kresse, 1999Kresse}. The Perdew-Burke-Ernzerhof (PBE) functional \cite{1996Perdew} was used to describe the exchange-correlation energy within the generalized gradient approximation (GGA). The cutoff kinetic energy was set to 350\,eV. Eight-layer 6$\times$6 supercell of the Si(100) surface with a 14\,{\AA} vacuum layer was used. P and Br atoms were placed on the top Si(100) surface, which was reconstructed to form dimers with one Si up and the other Si down. The dangling bonds at the bottom Si surface where saturated with hydrogen. During optimization, the bottom three Si layers were fixed, while the other atoms were fully relaxed until the residual forces were smaller than 0.01\,eV/\,{\AA}. A 3$\times$3$\times$1 k-point grid was used in the calculations. STM images were generated within the Tersoff-Hamann approximation \cite{1985Tersoff}. Taking into account the buckling of Si dimers, STM images were averaged over two different dimer slopes. To calculate the activation barriers, the nudged elastic band (NEB) method \cite{1998NEB} was employed using a 4$\times$4 supercell with structure c(4x2) and a convergence threshold for residual forces of 0.03\,eV/\,{\AA}.

\section{Results and Discussion}
The PBr$_3$ adsorption on the Si(100) surface was studied in detail in our previous work \cite{2023Shevlyuga}.
As it was shown, the PBr$_3$ molecule completely dissociates after adsorption on the Si(100) surface at 300\,K \cite{2023Shevlyuga}. At 77\,K, we also observed complete dissociation of the molecule. After PBr$_3$ dissociative adsorption on Si(100), two surface structures were most frequently observed, which we designated S1 and S2 (Fig.~\ref{fig1}). Structures S1 and S2 differ only in the position of the P atom; the remaining three Br atoms are located identically on top of the Si atoms. In structure S1 (Fig.~\ref{fig1}a), phosphorus occupies the end-bridge position in the groove between the dimers, forming bonds with two Si atoms of neighboring dimers in the same row and a Si atom of the previous layer.  In structure S2 (Fig.~\ref{fig1}b), P atom occupies the bridge position within the silicon dimer. In experimental STM images, the S1 and S2 structures are clearly distinguishable (Fig.~\ref{fig1}). The end-bridge or bridge site of the P atom may be unambiguous identified since the top Si and Br atoms are visible. We would like to emphasis that the adsorption positions of atoms on Si(100) allow for straightforward identification of their types (P or Br). Since P occupies bridge sites while Br occupies positions on top of Si atoms (Fig.~\ref{fig1}), phosphorus and bromine can be distinguished in the STM image \cite{2023Shevlyuga}.

 In the most stable surface structure S1, the P atom is bonded to a Si dimer with a Br atom, forming a P-Si-Si-Br complex (Fig.~\ref{fig1}a).  The formation of such a complex is 0.45\,eV more favorable compared to when the P and Br atoms are far apart. Note that the close proximity of two Br atoms in one Si dimer has no effect on the P adsorption energy. The S2 structure with P in the bridge position (Fig.~\ref{fig1}b) is less favorable than S1 with P in the end-bridge position by 0.46\,eV. However, in the absence of Br, on the clean Si(100) surface, the end-bridge position of P is more stable than the bridge by only 0.18\,eV \cite{2009Bennett} . Thus, the formation of the P-Si-Si-Br complex additionally stabilizes phosphorus in the end-bridge position. Nevertheless, formation of the S2 structure is also favorable for Br and P since moving P to a different row from the Br atoms reduces the adsorption energy by 0.12 eV.

We observed several diffusion paths for phosphorus during scanning at 77\,K and 300\,K. Below, we indicate the temperature at which each diffusion path was most frequently observed. However, we do not separate the diffusion paths at different temperatures into distinct sections, as some paths were observed at both temperatures. Figures~\ref{fig2}a--c shows phosphorus diffusion between the two most favorable positions, end-bridge and bridge. The transformation from structure S1 to S2 was reversible. The break in the line scan above the P atom in Fig.~\ref{fig2}b indicates that the P transition occurred when the tip passed over the atom. This suggests the impact of scanning on the P diffusion mechanism similar to the diffusion of hydrogen \cite{2000Stokbro} and bromine \cite{2022PavlovaJChemPhys} inside the dimer on Si(100), where line scan breaks were also detected above the atom. In the presence of three Br atoms, the activation barrier for P diffusion between the end-bridge and bridge sites is 1.18\,eV (Fig.~\ref{fig2}d). This diffusion path with and without an adsorbate on Si(100) has been calculated in previous studies, yielding comparable barriers. In the case of three H atoms located in place of Br, the activation barrier for P diffusion is 1.00\,eV \cite{2016Warschkow}, 0.94\,eV on a clean silicon surface \cite{2009Bennett}, and 1.28\,eV on a chlorinated surface with three Cl vacancies near the P atom \cite{Pavlova2025JCP}.

Phosphorus diffusion between the end-bridge and bridge positions was frequently observed at 77\,K, but very rarely at 300\,K. We would like to emphasize that at 77\,K we observed P in both the end-bridge and bridge positions, whereas at 300\,K we mostly observed P in the end-bridge and rarely in the bridge position. This difference can be explained by the presence of the energy barrier of 0.72\,eV (Fig.~\ref{fig2}d) between the bridge and end-bridge positions which can be difficult to overcome at 77\,K but possible at 300\,K. Thus, as will be shown below, we mostly observed P diffusion between end-bridge positions at 300\,K, passing through the bridge position as a local minimum. At 77\,K, phosphorus predominantly diffuses from the bridge position to the end-bridge or the bridge position on the adjacent dimer.

In the next diffusion path, the P atom moves between two end-bridge positions on the opposite sides of a dimer row (Fig.~\ref{fig3}a,b). In this case, one Br atom diffuses to a Si atom of the same dimer, resulting in the identical most stable structure S1. We calculated the minimum energy path and found that it consists of three elementary pathways (Fig.~\ref{fig3}c). Initially, phosphorus moves to the bridge position from the end-bridge position with a barrier of 1.18\,eV, as in the diffusion path discussed above. Then, bromine goes to the nearby Si atom of the same dimer across a 0.44\,eV barrier. Finally, phosphorus moves from the bridge to the end-bridge position on the other side of the dimer row, symmetrically to the first elementary pathway. The resulting diffusion barrier was found to be 1.18\,eV, as in the previous diffusion path (Fig.~\ref{fig2}). However, in contrast to the previous diffusion path, this diffusion pathway was observed predominantly at room temperature since P mostly locates in the end-bridge site at 300\,K.

We also observed P diffusion between end-bridge positions on one side of the dimer row (Fig.~\ref{fig4}a,b).
To form the most favorable structure S1 in the final state, one Br atom has moved toward the P atom.
We calculated two diffusion pathways for the phosphorus atom when it moves inside the groove and dimer row. The lowest barrier of 1.81\,eV was obtained when the P atom moved within the dimer row through the bridge position on the central dimer (Fig.~\ref{fig4}c). The relatively high activation barrier can be the reason why we very rarely observed this diffusion path, and only at room temperature.

\begin{figure}[h]
\begin{center}
\includegraphics[width=0.8\linewidth]{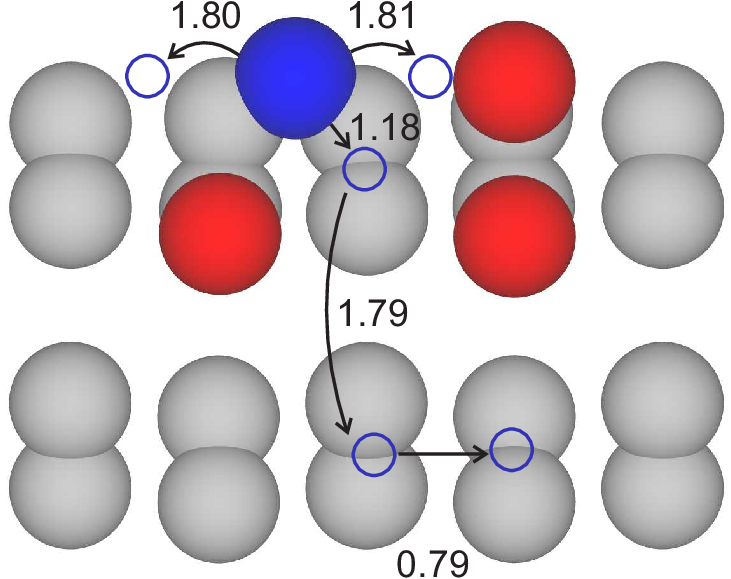}
\caption{\label{fig8}Diffusion paths of phosphorus initially located in the S1 structure on Si(100). Activation barrier values are given in electronvolts. Si atoms are shown in gray, Br atoms in red, and the P atom in blue. The final state is indicated by an open circle.}
\end{center}
\end{figure}

As mentioned before, we observed phosphorous diffusion between bridge sites at 77\,K. Figure~\ref{fig5} shows a series of STM images of the phosphorus diffusion from the bridge position in structure S2 to the bridge positions in the adjacent dimer row. The transition of phosphorus from the bridge position (Fig.~\ref{fig5}a) to the adjacent dimer row occurred while scanning the second frame (Fig.~\ref{fig5}b). Consequently, in the next frames (Fig.~\ref{fig5}c,d), phosphorus is no longer present in the bridge position within structure S2. The third frame shows a part of P atom in the neighboring dimer row, followed by numerous line breaks while scanning downwards. Upon further scanning, only such short line breaks within one dimer row are visible in the STM image (Fig.~\ref{fig5}d). Although we have no evidence that the line breaks are caused by the diffusion of the P atom, we can suggest that phosphorus moves quickly during scanning since there are no Br atoms in this row to block its movement. We calculated the activation barriers for P diffusion to the adjacent row and within the row. The barrier of the P transition between the bridge positions of the neighboring dimer rows is quite high, 1.79\,eV (Fig.~\ref{fig5}e). Thereby we believe that this transition occurs under the influence of the tip, particularly since diffusion was observed when the tip was above the P atom (Fig.~\ref{fig5}b). The P diffusion pathway within the dimer row has an energy barrier of 0.79\,eV (Fig.~\ref{fig5}f). This is the lowest barrier among all calculated pathways for phosphorus diffusion.

Additionally, we observed phosphorus diffusion toward the defect, which is an adsorbed oxygen atom that appeared on the surface due to water adsorption. Figure~\ref{fig6}a shows a completely dissociated PBr$_3$ molecule, a dissociated water molecule (OH and H) \cite{2015Smith} and an oxygen atom. We identified the oxygen atom based on the following two facts. First, another similar defect formed on the surface at the location of the OH (Fig.~\ref{fig6}a,b). Second, simulated STM image of the O atom between the Si atoms of the dimer, averaged taking into account Si buckling (Fig.~\ref{fig7}a,b), is similar to the experimental STM image of the defect (Fig.~\ref{fig6}a,b). Initially, phosphorus was in the end-bridge position (Fig.~\ref{fig6}a) and then diffused to the end-bridge position on the opposite side of the dimer row (Fig.~\ref{fig6}b), similar to the diffusion shown in Fig.~\ref{fig3}. Upon scanning, P moved to a position near oxygen (Fig.~\ref{fig6}c). With further scanning, phosphorus hoped to another oxygen atom (Fig.~\ref{fig6}d) and back (Fig.~\ref{fig6}e,f) through several dimer rows. We support the observation of P and O co-adsorption with calculations of adsorption energies. We calculated the adsorption of P in the end-bridge site several dimers distant from the oxygen atom in the nearby dimer row (Fig.~\ref{fig7}a,b) and near the oxidized dimer in the observed configuration (Fig.~\ref{fig7}c,d). Indeed, the phosphorus end-bridge position near the oxidized dimer is 0.1\,eV more favorable than far from it. The interaction of P and O atoms occurs through one common Si atom with the formation of the Si-P-Si-O-Si complex(Fig.~\ref{fig7}c). Thus, in the presence of oxygen, which appears on the active Si(100) surface as a result of water adsorption, phosphorus prefers to be located near the oxidized Si dimer.

Note that the activation barriers were estimated without considering additional parameters such as tip effect or phosphorus charge. However, such parameters may affect diffusion. In particular, P hops when the tip passes over it (see Fig.~\ref{fig2}b, Fig.~\ref{fig5}b, and Fig.~\ref{fig6}e), supporting the suggestion that the tip had an impact. Charging of surface structures was reported for various fragments of the PH$_3$ molecule on the Si(100) surface \cite{2016Warschkow}. Also, current and voltage may alter diffusion, as was demonstrated for hydrogen \cite{2000Stokbro} and bromine \cite{2022PavlovaJChemPhys} hops within one dimer.

Figure~\ref{fig8} shows a summary of the considered diffusion paths of phosphorus initially located in the S1 surface structure. For completeness, we additionally calculated the diffusion of phosphorus from the end-bridge position to the left to the same position (path with a barrier of 1.80\,eV), although we did not observe it. Bromine inhibits phosphorus diffusion across the Si(100) surface for two reasons. First, bromine stabilizes the location of phosphorus in the S1 and S2 structures, especially in the S1 structure due to the formation of the P-Si-Si-Br complex. Second, upon diffusion, phosphorus preferentially forms bonds with Si, which is not bonded to Br, i.e., has a dangling bond. Indeed, phosphorus diffusion inside the dimer row without bromine has the lowest activation energy, around 0.8\,eV. However, to move to a bromine-free surface area, phosphorus must first overcome a barrier of around 1.8\,eV (Fig.~\ref{fig8}). Thus, the presence of three Br atoms near a P atom blocks the P diffusion path with the lowest barrier, within the dimer row. Note that P and Br may not be located on adjacent dimers after PBr$_3$  dissociative adsorption; in this scenario, bromine will no longer restrict phosphorus diffusion.

For atomically precise doping of silicon, adsorption of a phosphorus-containing molecule is followed by exchange of the P atom with the Si atom, which occurs upon heating \cite{2003Schofield}. It is critical to minimize the displacement of the P atom from its precisely defined position before incorporation. Limiting P diffusion with Br atoms facilitates atomically precise incorporation. Considering the diffusion of phosphorus surrounded by bromine at the lowest temperature studied, 77\,K, its diffusion with a minimum barrier of 1.18\,eV occurs between the end-bridge and bridge positions (Fig.~\ref{fig2}). According to previous calculations \cite{Pavlova2025JCP}, phosphorus has a minimal energy path of exchange with the same silicon atom from both the end-bridge and bridge positions. As a result, such diffusion does not affect the final position of P-Si exchange. At 300\,K, phosphorus can already diffuse between end-bridge positions on opposite sides of the dimer row, accompanied by bromine diffusion (Fig.~\ref{fig3}). In this case, a change in the positions of phosphorus and bromine causes a shift in the preferred site of P insertion onto the neighboring Si atom \cite{Pavlova2025JCP}.

\section{Conclusions}

The diffusion pathways of phosphorus on the Si(100) surface have been studied at 77 and 300\,K after dissociative adsorption of the \ce{PBr3} molecule. At 77\,K, the bridge site is quite stable for phosphorus, whereas at 300\,K, phosphorus prefers to move to a more stable end-bridge site, overcoming the energy barrier. Consequently, diffusion between bridge positions or between the bridge and end-bridge positions was mostly observed at 77\,K. Instead, P diffusion between end-bridge sites on opposite sides and along one side of the dimer row was observed at 300\,K. In this case, P diffusion was accompanied by the diffusion of a Br atom in order to preserve the most favorable configuration of P and Br atoms on the surface. The phosphorus diffusion is hindered by bromine, according to calculated activation barriers of different diffusion pathways. Additionally, it was shown that phosphorus position is more stable next to oxidized dimer, which forms on the silicon surface due to water adsorption, than on the clean surface. The obtained results significantly expand the understanding of phosphorus diffusion on the Si(100) surface, which is particularly relevant for phosphorus incorporation into the Si(100) surface with atomic precision.

\begin{acknowledgments}
This research was supported in part through computational resources of HPC facilities at HSE University.
\end{acknowledgments}

\bibliography{paper_P_dif_arxiv}

\end{document}